\documentclass[doublecol]{epl2}
\usepackage{morefloats}
\usepackage{amsmath}
\usepackage{amsfonts}
\usepackage{graphicx}
\usepackage{dcolumn}
\usepackage{bm}
\usepackage{color}

\usepackage{hyperref}
\title{Synchronization in Delayed Multiplex Networks}
\author{$^1$Aradhana Singh, $^1$Saptarshi Ghosh, $^{1, 2}$Sarika Jalan\footnote{$^*$Corresponding author: sarikajalan9@gmail.com}, and $^{3, 4}$J\"urgen Kurths}
\institute{ \inst{1} Complex Systems Lab, Discipline of Physics, Indian Institute of Technology Indore, Khandwa Road, Indore-452017, India \\
 \inst{2} Centre for Biosciences and Biomedical Engineering, 
Indian Institute of Technology Indore, Khandwa Road, Indore-452017, India \\ 
\inst{3} Potsdam Institute for Climate Impact Research, P.O. Box 601203, D-14412 Potsdam, \\
\inst{4} Institute 
for Complex Systems and Mathematical Biology, University of Aberdeen,
Aberdeen-AB243FX \\}
\abstract
{
 We study impact of multiplexing on 
the global phase synchronizability of different layers in the delayed coupled multiplex networks.
We find that at strong couplings, the multiplexing induces the global synchronization in sparse networks. 
The introduction of global synchrony depends on the connection density of the layers 
being multiplexed, which further depends on the underlying network architecture. 
Moreover, multiplexing may lead to a transition from a quasi-periodic or chaotic evolution to a periodic evolution.  For the periodic case, the multiplexing 
may lead to a change in the period of the dynamical evolution.
Additionally, delay in the couplings may bring upon synchrony to those multiplex networks which do not exhibit synchronization for the undelayed evolution. Using a simple example of two globally connected layers forming a multiplex network, we show how delay brings upon a possibility for the inter layer global synchrony, that is not possible  for the undelayed evolution.
}

\begin{document}
\maketitle

{\bf Introduction:} Realization that many real-world systems such as; transport, banks, stock market 
\cite{Multiplex_Arenas1,Multiplex_Arenas2, Multiplex_degree, home_2012,wasserman_1994} etc 
can be represented by multiple levels of interactions, has lead to a spurt in 
the activities of understanding and characterizing various properties
of multiplex networks. The prime motivation of the multiplex
framework is that the function of individuals in one level get affected by the interactions and functions in the other levels. 
A Multiplex network consists of layered network with one to one correlation between the replica
 nodes in different layers \cite{two_layer_lattice_map, two_layer_lattice_osc1, two_layer_lattice_osc2, bocc_2014,lee_2012,Multi_layer}. Each layer in the multiplex network constitutes different 
types of relations between the same units, presenting a more realistic  framework 
of modeling real world interactions \cite{Multiplex_Arenas1}.
Further, one of the most fascinating emergent behavior of interacting 
 nonlinear dynamical units is the observation of synchronization \cite{ulrike,pil_2008,roy}, 
which is defined as the appearance of a relation 
between two processes due to the interactions 
between them \cite{SJ_2003,kur_1984}. Synchronization has 
been investigated a lot due to its wide range of  applicability \cite{pil_2008,kur_1984}.
For example, synchronization plays a crucial role in proper functioning 
of systems as diverse as motor functions of a neural 
network \cite{laird_2001}, efficiency in a business or 
academic system \cite{Multiplex_Arenas2}, signal processing in a communication network to proper 
flow of traffic in transport networks \cite{transpot_multiplex}.
A recent work has revealed that there exists an onset of explosive 
synchronization in multilayer networks \cite{Multiplex_explosive,Boccaletti_explosive}. 
Other recent works have investigated changes in the static and 
dynamic behavior of multiplex networks with the interlink strength variation \cite{Multiplex_interlink} as well as have reveled the intra-layer synchronization without the inter-layer synchronization \cite{Inra_syn_multiplex}.  

Furthermore, delays naturally arise in real world systems due to the finite speed of information propagation  \cite{lask_2010}. The delays are shown to lead to many emerging phenomena in coupled dynamical units
such as oscillation death, stabilizing periodic orbits,
enhancement or suppression of synchronization, chimera state, etc
\cite{osc_death_delay,Atay_analytical,Thilo,delay_enhance_syn,Pre_rapid,delay_coup_osc,
delay_suppress_syn,chimera, wang_delay, neuron_delay}. 
In this Letter, we investigate impact of multiplexing as well as delay on the global synchronization of various networks.
 Instead of exact synchronization we analyze the
phase synchronization as sparse networks exhibit a negligible number of nodes manifesting the exact synchronization.
Furthermore, in many realistic situations the connection density as well as degree 
distribution of various layers can be different. 
For instance, a collaboration and a friendship networks 
formed by the people
employed in the same institution 
can have different architecture as well as connection density, and hence here we consider different layers 
being represented by different network architectures.
\begin{figure}[ht]
\centerline{\includegraphics[width=3.0in, height=1.4in]{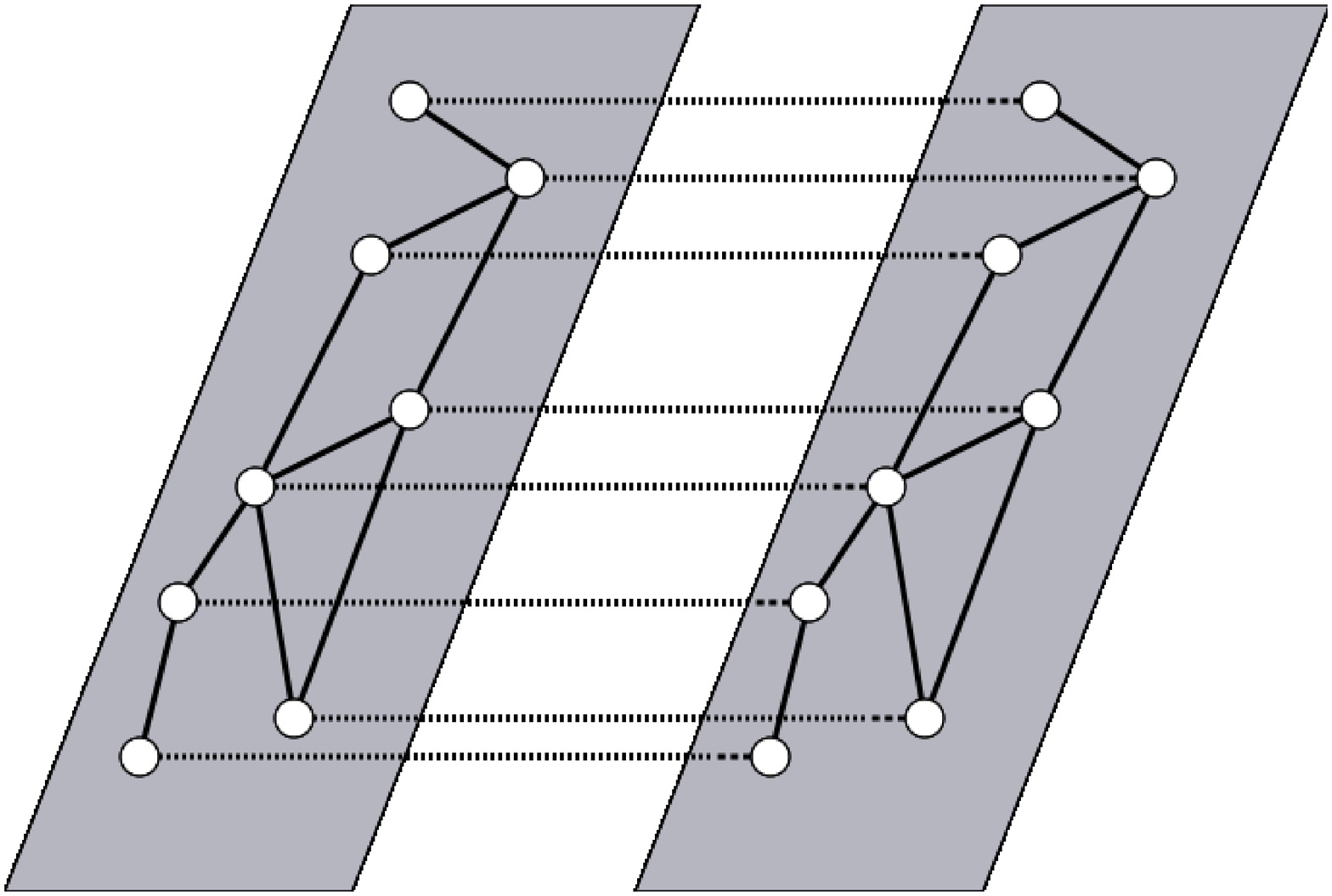}}
\caption{Schematic diagram depicting a two layer multiplex network. Solid lines represent the 
intra-layer connections and the dashed lines represent the inter-layer connections.}
\label{Fig_multiplex}
\end{figure}

\begin{figure}
\centerline{\includegraphics[width=3.5in, height=2.0in]{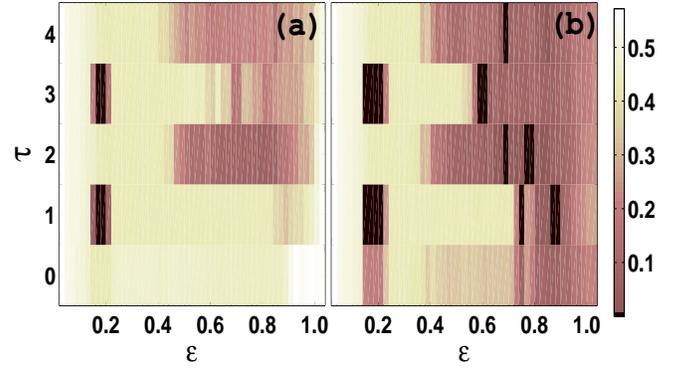}}
\caption{Phase diagram depicting the variation of $D$ for SF network of $\langle k \rangle = 4$(a) when it is isolated and (b) after multiplexing with another SF network of 
 $\langle k \rangle = 10$, with respect to $\varepsilon$. The number of nodes in each layer is $N=250$. All results are averaged over $20$ different realization.
}
\label{Fig_SF4}
\end{figure}
{\bf Theoretical framework:} We consider a network of $N$ nodes and $N_c$ connections.
 The dynamical 
evolution of each node at time $t$ in the network is represented by a 
variable $x^i(t), i=1,2,......,N$.  
This evolution of the dynamic variable with the delay ($\tau$) can be described by 
a delayed coupled map model as \cite{newman_2010};
\begin{equation}
x_i(t+1)=(1-\epsilon)f(x_i(t))+\frac{\epsilon}{k_i}\sum_{j=1}^{mN}A_{ij}
f(x_i(t-\tau))
\label{cml}
\end{equation}
where $\varepsilon$ is the overall coupling strength ($0\leq \epsilon\leq 1$), 
$\tau$ represents the communication delay
between the nodes, and $m$ is the number of layers in the multiplex network.  
$A$ is the adjacency matrix with elements $A_{ij}$ taking values 1 and 0 
depending upon whether there exists a connection between nodes $i$ and $j$ or not. 
We consider undirected network with no self loop and hence  $A$ is a symmetric matrix with diagonal elements being zero.
Without loss of generality, we consider a simple multiplex network with two layers and $N$ nodes in each layer, adjacency matrix $A$ for a two layer multiplex network can be given as;
\begin{equation}
    A=
      \begin{pmatrix} A^1 & I \\ I & A^2 \end{pmatrix}, 
\end{equation}
where, $A^1$ and $A^2$ are the adjacency matrices corresponding to the layer 
$1$ and layer $2$.   
$k_i=(\sum_{j=1}^{N} A^l_{ij})+1$ is the degree of the i-th node in the $l^{th}$ 
layer of the multiplex network.
The function $f(x)$ defines the local nonlinear map and the coupling between the nodes. 
In the present framework, 
we consider local dynamics given by the 
logistic map; $f(x) = 4 x(1-x)$ and the circle map, for which $f(x)$ shows chaotic evolution \cite{strogratz_2001}.

\begin{figure}[ht]
\centerline{\includegraphics[width=3.4in, height=1.8in]{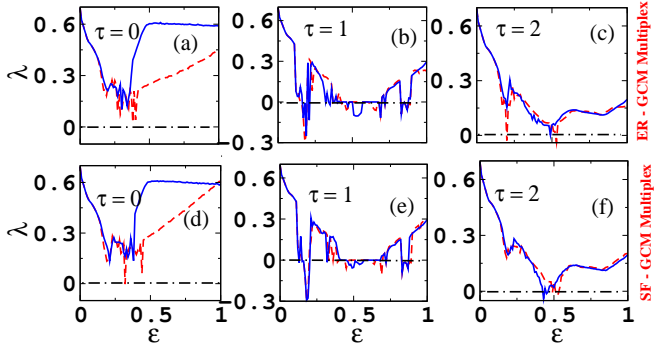}}
\caption{Variation of the largest Lyapunov exponent with 
$\varepsilon$ for the isolated SF network with $\langle k_1 \rangle = 10$ (dotted line) and after multiplexing with the globally connected network (solid line) for (a) undelayed, 
(b) delayed ($\tau=1$), (c) delayed ($\tau=2$) 
 evolution. Similarly, for isolated random networks with $\langle k \rangle = 10$ (dotted lines) and after multiplexing  with the
globally connected networks (solid line) for (d) undelayed, (e) delayed ($\tau=1$), 
(f) delayed ($\tau=2$) 
 evolution. 
Here number of nodes $N$ is taken to be $50$ in each layer. 
All plots are for average over 20 different realizations of the initial conditions.}
\label{Fig_lya1}
\end{figure}

We quantify global phase synchronization defined as follows \cite{SJ_2003}. 
Let $n_i$ and $n_j$ denote the number of times when the variables 
$x_i(t)$ and $x_j(t)$, $t= 1,2,......T$ for the nodes $i$ and $j$ 
exhibit local minima during the time interval T. Let $n_{ij}$ denotes the number of times these local minima matches each other. The phase distance between two nodes $i$ and $j$ is then given as 
$$
d_{ij}=1-\frac{2n_{ij}}{(n_i+n_j)}
$$
The nodes i and j are phase synchronized if 
$d_{ij}=d_{ji}=0$. We used as global phase 
synchronization measure $D=\sum_{j=1}^{nN} d_{ij}$ for the whole multiplex 
network, $D_{1}=\sum_{j=1}^{N} d_{ij}$ and $D_{2}=\sum_{j=N}^{2N} d_{ij}$ for the first and the second layer respectively. The global phase synchronized state 
for the whole multiplex network exists for $D = 0$,  whereas one of the layers being global 
synchronized  is indicated by $D_{1} = 0$ or $D_{2} = 0$ and $D\neq 0$.
We study global phase synchronizability of a 1-d lattice, scalefree (SF), random 
and the globally connected networks upon multiplexing with another layer 
of 1-d lattice, SF, random 
and globally connected network architectures. 
The 1-d lattices used in the simulation have 
circular boundary conditions
with each node having $k$ nearest neighbors. SF and random networks
are obtained by using BA and ER models respectively
\cite{rev-network}. The multiplex network is constructed by making one to one 
connections between the replica nodes in two layers. 

{\bf Results:}
We evolve Eq.~\ref{cml} starting from a set of random initial 
conditions and study global phase synchronization after an initial transient for various combinations such as regular-regular, regular-random, random-global, SF-global, random-random and SF-SF multiplex networks.
First, we discuss changes in global phase synchronization
of sparse SF networks upon multiplexing with a SF network. 
We take SF network ($\langle k_1 \rangle = 4$) and multiplex it with 
another SF network  ($\langle k_2 \rangle = 10$). 
For the undelayed evolution, an isolated SF network with 
$\langle k_1 \rangle = 4$ does not 
exhibit global phase synchronization for any of the 
coupling values as depicted in Fig.~\ref{Fig_SF4}(a). 
The multiplexing in this case does not 
bring upon any change in global synchronizability.

An interesting phenomenon is displayed for the delayed evolution, while at the weak coupling multiplexing does not influence the synchronizability,  at strong couplings multiplexing brings upon the global synchronization in the 
 layer having sparse connections. The isolated SF networks, irrespective of the average
 degree of the network, exhibit the global phase synchronization at weak coupling for the
 odd parity of delays in the coupling range $0.16\overset{\sim}{<} \varepsilon \overset{\sim}{<} 0.18$ (Fig.\ref{Fig_SF4}). The dynamical evolution
in this coupling range is periodic with periodicity depending on the delay value (Fig.~\ref{Fig_lya1} (b)). 
While the multiplexing does not introduce 
any significant change in the synchronizability of the network
for this coupling 
range, it leads to an enhancement in the global phase synchronizability at strong couplings. 
Thus, the delayed coupled isolated network ($\langle k_1 \rangle = 4$) does not exhibit a
 global synchronization, and  multiplexing with different denser networks induces the global synchrony.
\begin{figure}[ht]
\centerline{\includegraphics[width=3.5in, height=1.6in]{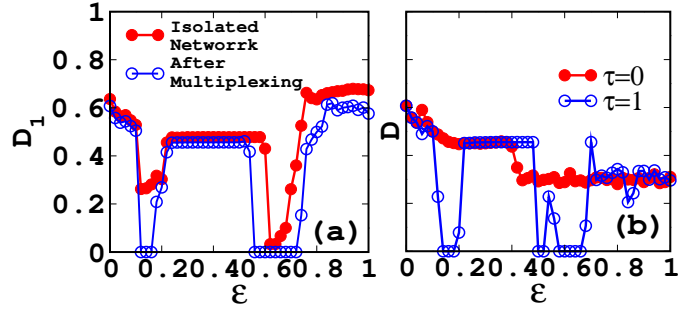}}
\caption{(a) Variation of $D_{1}$ for the 1-d lattice ($\langle k_1 \rangle=10$)
before (closed circles) and after multiplexing (open circles) with the globally connected network for $\tau =1$.
(b)  Variation of $D$  for a multiplex network consisting of two globally connected layers for 
the undelayed (closed circles) and for the delayed evolution (open circles).
Number of nodes in each layer is $N=250$. 
}
\label{Fig_glob_NNC10}
\end{figure}
The coupling range for which the global synchronization is induced may change with the delay 
value even when the network architecture remains the same. For example at $\tau=1$ the global
 synchronization is induced for the coupling range $ 0.71\overset{\sim}{<} \varepsilon \overset{\sim}{<} 0.72 $ and $ 0.84\overset{\sim}{<} \varepsilon \overset{\sim}{<} 0.86 $, and for $\tau=3$ 
 synchronization is observed for the coupling range $ 0.56\overset{\sim}{<} \varepsilon \overset{\sim}{<} 0.58 $. 
The same phenomenon is also observed for the other network architectures which in the isolated
state do not exhibit the synchrony for the delayed evolution but upon multiplexing with
 the another denser network exhibit the global synchronization at strong coupling values. 
One such example is sparse 1-d lattice, which in the isolated state does not exhibit the
 global 
synchronization at strong couplings, but upon multiplexed with the globally connected
 network exhibit the global synchronization in the coupling 
range $ 0.56\overset{\sim}{<} \varepsilon \overset{\sim}{<} 0.76 $ (Fig.~\ref{Fig_glob_NNC10}(a)).

\begin{figure}[ht]
\centerline{\includegraphics[width=3.5in, height=2.0in]{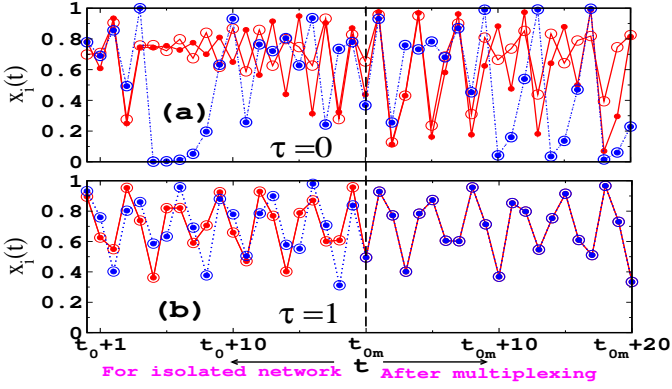}}
\caption{Time series of four nodes (open and closed circles) from the globally connected layers (two from each) before and after multiplexing for  (a) undelayed  and (b) delayed ($\tau=1$) evolution. $N=250$ in each layer and $\varepsilon=0.65$. The time series is potted after the initial transient of 10,000 time steps before ($t_0$) and after multiplexing ($t_{0m}$).
 }
\label{Glob_glob_time}
\end{figure}

Further, two layers do not get synchronized with each other for the undelayed evolution, whereas  introduction of the delay induces the global synchronization in the multiplex network. 
 An introduction of delay is already known to enhance the synchronizability of a network \cite{Atay_prl}, however, the finding that the multiplex 
network exhibit the global synchronization for the delayed evolution is more 
interesting  as for the undelayed evolution the multiplex networks does not
exhibit the global synchronization even if the globally connected network
form the individual layer (Fig.~\ref{Fig_glob_NNC10}(b)).
Note that the isolated networks for sufficiently high connection density are
known to exhibit the global synchronization for the undelayed evolution. 
\begin{figure}[ht]
\centerline{\includegraphics[width=3.5in, height=2.0in]{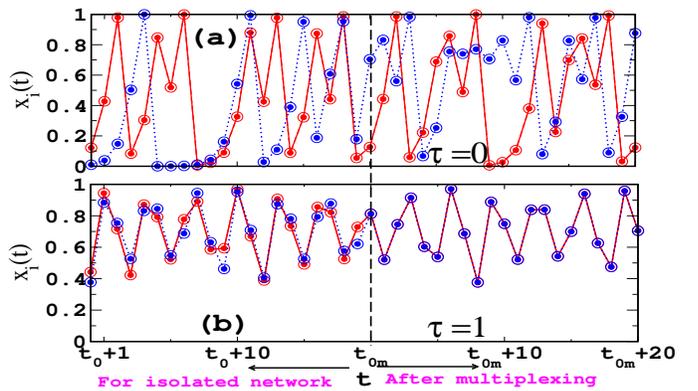}}
\caption{Time series of two nodes (open and closed circles with solid lines) from 
the SF network (layer 1) and two nodes (open and closed circles with dotted lines) from
the globally connected network (layer 2) before and after multiplexing for (a) undelayed  and (b) delayed ($\tau=1$) evolution. $N=250$ in each layer and $\varepsilon=0.69$. For the SF 
network, $\langle k_1 \rangle = 10$. The time series is potted after the initial transient of 10,000 time steps before ($t_0$) and after the multiplexing ($t_{0m}$).
}
\label{SF10_glob_time}
\end{figure}
For example, the multiplex network consisting of the SF layer with  $\langle k_1 \rangle = 10$ and the globally connected layer does not 
exhibit the global synchronization at $\varepsilon=0.69$ for the undelayed evolution (Fig. \ref{SF10_glob_time}(a)).
Whereas, the introduction of delay results in the global synchronization of the multiplex network (Fig. \ref{SF10_glob_time}(b)).  
For the case of a multiplex network consisting of two globally connected 
layers, the undelayed evolution does not bring upon the global synchronization 
to the whole network, while the individual layer keeps on showing the global synchronization as observed for the isolated globally connected network (Fig. \ref{Glob_glob_time}(a)). An introduction of the delay leads to the global synchronization of the multiplex network (Fig. \ref{Glob_glob_time}(b)).
Similarly, an introduction of the delay causes global synchronization in the multiplex network consisting of the random and the globally connected layers, which was
 not observed for the undelayed evolution as discussed above.

The introduction of global synchrony for the delayed evolution can be explained by considering the case of exact synchronization in a simple network architecture as follows.   Let $x_i^1(t) = y^1(t), \,\, \forall i \text{and} \forall t>t_0$ be the global synchronized state of a globally connected isolated network and $z_i^1(t) = y^2(t)$ be the global synchronized state of another isolated network. 
Upon multiplexing,  the difference variable between the
two nodes in the same layer $2$ at $\varepsilon=1$ will be given as;
\begin{eqnarray}
dx^2_{ij}(t+1) = \frac{(k^1_i-k^1_j)}{(k^1_i+1)(k^1_j+1)} ( f(y^1(t-\tau)) - f(y^2(t-\tau)));
 \nonumber\\
\end{eqnarray}
Thus, for $k^1_i=k^1_j$, which is the case of the globally connected network, the intra-layer synchronization manifested by the isolated networks remains unaffected after multiplexing. 
Whereas, the difference variable between the mirror nodes $i$ and $j$ from two different layers will be;
\begin{eqnarray}
x_i^1(t+1) - x_i^2(t+1) = f(y^1(t))-f(y^2(t))
 \nonumber
\end{eqnarray}
The above difference variable will not vanish for the  
nodes having the chaotic dynamics and therefore restricting the synchronization 
between them. 
However, for the delayed evolution the difference variable will be;
\begin{eqnarray}
x_i^1(t+1) - x_i^2(t+1) = (1-\varepsilon) f(y^1(t))-f(y^2(t)+ 
 \nonumber\\ (\varepsilon)( f(y^1(t-\tau)) - f(y^2(t-\tau)));
\end{eqnarray}
and depending on the value of $\varepsilon$ and $\tau$, nodes from the different layers may get synchronized even when both nodes have chaotic dynamics.

\begin{figure}[t]
\centerline{\includegraphics[width=3.5in, height=1.5in]{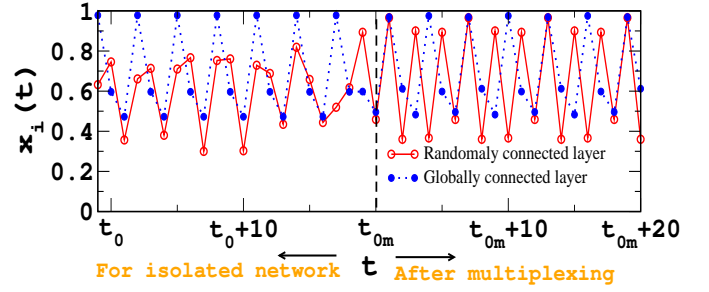}}
\caption{Time series  of two nodes (open and closed circles) from the
random network (layer 1) and two nodes (open and closed circles) from the globally connected network (layer 2) before and after multiplexing. $N=50$ in each layer, $\varepsilon=0.58$ and for the random network has average degree $\langle k_{1} \rangle = 10$. The time series is potted after the initial transient of $10,000$ time steps before ($t_0$) and after multiplexing ($t_{0m}$).
}
\label{Fig_timeseries}
\end{figure}
Further, in order to investigate the changes in the dynamical evolution upon 
multiplexing, we calculate the largest Lyapunov exponent as a function of $\varepsilon$ (Fig. \ref{Fig_lya1}). 
For the undelayed evolution, the multiplexing does not bring upon any significant change in the dynamical evolution and the dynamics remains chaotic for all the coupling 
values as observed for the isolated random network  (Fig.\ref{Fig_lya1} (a) ).
Whereas, for the delayed evolution,
 multiplexing  may lead to  a transition from the quasi-periodic to periodic, or from the chaotic to a periodic evolution.  
For example, in the coupling range $0.54 \overset{\sim}{<} \varepsilon \overset{\sim}{<} 0.58$  and for $\tau=1$, the multiplexing leads to a transition from the quasi-periodic to a periodic evolution (Fig.\ref{Fig_lya1} (b)). 
In the same coupling range an isolated network exhibit the global phase synchronization, 
while multiplexing destroys the global phase synchronization as discussed above. 
In the coupling range 
($0.83 \overset{\sim}{<} \varepsilon \overset{\sim}{<} 0.87$), where the isolated network leads to
a periodic evolution, multiplexing retains the periodic evolution with the same period (Fig.\ref{Fig_lya1} (b)).
But for the same coupling range the multiplexing also retains the global phase 
synchronization manifested by the isolated random network. 
In the coupling range $0.88 \overset{\sim}{<} \varepsilon \overset{\sim}{<} 0.89$, 
the multiplexing leads to a transition from a periodic to the chaotic evolution 
(Fig.\ref{Fig_lya1} (b)).  

\begin{figure}[ht]
\centerline{\includegraphics[width=3.5in, height=1.4in]{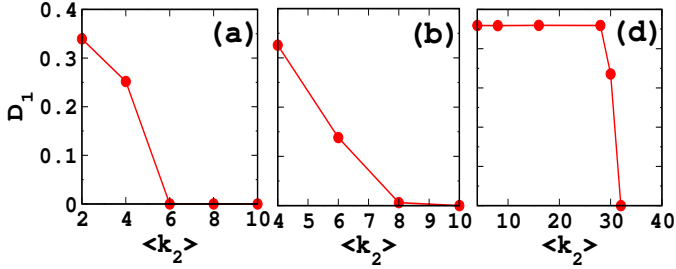}}
\caption{Variation of ($D_{1}$) for a SF network with $\langle k_1 \rangle = 4$ multiplexed with 
(a) SF network at $\varepsilon=0.86$, (b) random network at $\varepsilon=0.86$ and (d) 1-d lattice at $\varepsilon=0.76$,  for 
various average degrees.
Here number of nodes $N$ is taken to be $250$ in each layer. 
All results are for average over $20$ different realizations of the initial condition.}
\label{Fig_2}
\end{figure}

Furthermore, 
as discussed above, the globally coupled network exhibits 
a transition from a periodic state to another periodic state 
with a different period upon multiplexing. For example, in the coupling range 
$0.54 \overset{\sim}{<} \varepsilon \overset{\sim}{<} 0.58$, the isolated random network for $\tau=1$ exhibits the
chaotic dynamics, and the isolated globally connected network 
shows a periodic evolution with the periodicity three. The multiplexing leads to a periodic state with 
periodicity six for both the networks (Fig.\ref{Fig_timeseries}). 
 Note that in the same coupling range there is no synchronization between nodes of 
the different layers (Fig.\ref{Fig_timeseries}). 

Furthermore, in order to see an impact of the network architecture on the enhancement 
 in the global synchronizability of a network upon multiplexing, 
we present results for the multiplexing of the SF network with various 
different network architectures viz SF, random network and 
1-d lattice of various average degrees.
We find that the network architecture plays an important 
role in deciding the denseness of connection in a layer which lead to the global synchronization. For instance, SF networks with $\langle k_1 \rangle =4 $ exhibit the global synchronization upon multiplexing with a SF network having  $\langle k_2 \rangle =6 $ (Fig.~\ref{Fig_2}(b), (c)), while the same phenomenon is observed for multiplexing with the random and the 1-d lattice of $\langle k_2 \rangle =10 $ and $\langle k_2 \rangle =30 $ respectively (Fig.~\ref{Fig_2}(b), (c)). 

In order to demonstrate the robustness of the phenomenon that the multiplexing introduces the global synchronization in the 
sparse networks at strong couplings, we present results for the coupled circle maps as well. The local dynamics then can be given by:
\begin{equation}
f(x) = x + \omega + (p/2\pi)sin(2\pi x) \;\; ~(mod 1).
\end{equation}
Here we discuss results with the parameters of circle map corresponding
to the chaotic evolution ($\omega = 0.44$ and $p = 6$).

\begin{figure}[t]
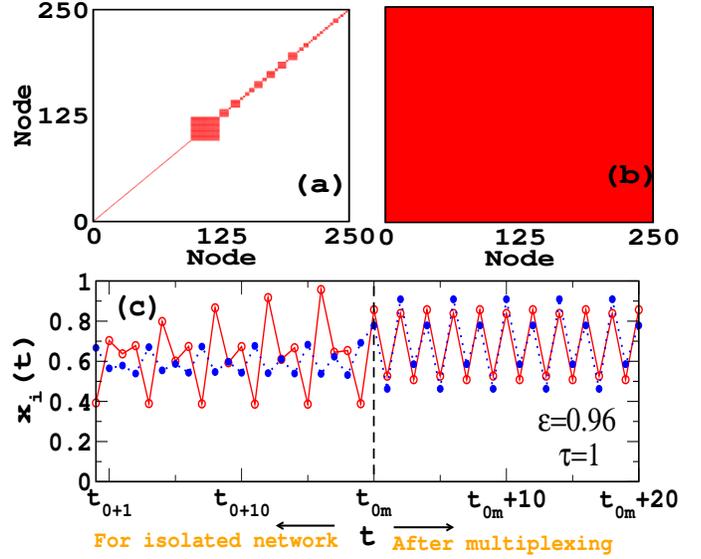

\centerline{\includegraphics[width=3.5in, height=1.4in]{Fig9a.eps}}
\centerline{\includegraphics[width=3.5in, height=1.5in]{Fig9b.eps}}
\caption{(a) and (b) node versus node diagram, (c) time series for SF networks with 
$\langle k_1 \rangle = 2$.  (a) Cluster state for the undelayed evolution, (b) globally synchronized state indicated by the formation of a single cluster after multiplexing with 
the another SF network having $\langle k_2 \rangle = 10$ and (c) time series of two nodes from
the SF network with $\langle k_1 \rangle = 2$ before and after the multiplexing.
Here number of nodes ($N$) is taken to be $250$ in each layer, $\varepsilon=0.96$ and $\tau=1$.}
\label{Fig_Node_Node}
\end{figure}

For the local evolution being governed by the circle map and the coupled dynamics given by 
Eq. \ref{cml}, the sparse networks, those do not exhibit the global synchrony, 
upon multiplexing with the denser networks exhibit the global phase synchronization.
Fig.~\ref{Fig_Node_Node} presents an example where the isolated SF network ($\langle k_1 \rangle =2 $) leads to the cluster formation (Fig.~\ref{Fig_Node_Node}(a)), while multiplexing 
with the SF network ($\langle k_2 \rangle =10 $) leads to a transition to the 
globally synchronized state at $\varepsilon=0.96$ (Fig.~\ref{Fig_Node_Node}(b)). 
 Furthermore, at the same coupling value there is a transition from the chaotic to the periodic evolution with periodicity four (Fig.~\ref{Fig_Node_Node}(b)). 

{\bf Conclusion:}
We have investigated an impact of multiplexing on the global phase synchronization and the dynamical 
evolution of the nodes in the individual layer in  delayed multiplex networks.
We mainly find that the impact of multiplexing depends on the network architecture of different layers as well as on the overall coupling strength. For the undelayed evolution and 
at the weak couplings for delayed evolution, multiplexing does not lead to any significant 
impact on the synchronizability of a network,  
yielding a similar dynamical evolution irrespective of the network architecture. For these cases, the same parity of delay values brings upon a similar impact as observed for the isolated networks \cite{Pre_rapid,Atay_prl}. 
Upon multiplexing, the odd parity of delay values exhibits the global phase synchronization with a periodic evolution, whereas in the same coupling range the even 
parity of delays may exhibit the global phase synchronization with the dynamical evolution being chaotic.  
At strong couplings, the delayed sparse networks exhibit the 
global synchrony upon multiplexing with the denser networks. The connection density of the
layer being multiplexed plays an important role in deciding if there will be an
introduction of the global synchronization in the sparse networks which
further depends on the network architecture.  Multiplexing with a SF network with the
connection density lower than than the corresponding ER and regular networks may
lead to the global synchrony in a sparse SF network at the strong couplings. 
Furthermore, for the undelayed evolution, the multiplex network does not display the global synchrony even though the nodes in each layer are globally connected while incorporation of delay leads to the global synchronization of the entire network. Using a simple network architecture
consisting of two globally connected layers, we demonstrate that for the chaotic intrinsic
dynamical evolution, the global synchronization in a multiplex network is not possible. The introduction of delays in the coupling provides a possibility for the same.
Additionally, the multiplexing leads to a transition from a quasi-periodic or the chaotic evolution to a periodic evolution as well as may lead to a change in the periodicity. 

This analysis can be extended further in getting a better understanding of various dynamical processes
 in real world systems, such as controlling congestion in the multiplex transport networks \cite{transpot_multiplex} as well as occurrence of different diseases such as epileptic seizure by representing brain as delayed multiplex networks \cite{brain_multiplex}.

\subsection{\bf Acknowledgement} SJ is grateful to DST (EMR/2014/00368)
 and CSIR (25(0205)/12/EMR-II) for financial support.  AS and SG thank complex systems lab members for useful discussions.


\end{document}